\documentclass[aps,twocolumn,pre,longbibliography,floatfix,superscriptaddress]{revtex4-1}
\usepackage{graphics,graphicx,subcaption,caption,lipsum}
\usepackage{amsfonts,amsmath,amssymb,physics,xcolor}
\usepackage{times}
\usepackage{hyperref}
\hypersetup{colorlinks=true,linkcolor=blue,citecolor=blue,urlcolor=blue}

\def\beqn{\begin{eqnarray}}
\def\eeqn{\end{eqnarray}}

\def\hw{\hbar \omega}
\def\hw4{ \frac {\hbar \omega}{4}}

\def\b0{b_0}

\newcommand{\ii}{\mathrm{i}}
\newcommand{\avg}[1]{\left\langle #1 \right\rangle}
\newcommand{\coloneqq}{\mathrel{:\mkern-0.25mu=}}

\newcommand{\Ord}[1]{\mathrm{O}\!\left(#1\right)}

\begin{document}

\title{Out-of-time-order correlators for Swanson Hamiltonian with interaction terms}
\author{M. W. AlMasri}
\email{mwalmasri2003@gmail.com}
\affiliation{Wilczek Quantum Center, School of Physics and Astronomy, Shanghai Jiao Tong University, Minhang, Shanghai, China}
\author{Marta Reboiro}
\email{reboiro@fisica.unlp.edu.ar}
\affiliation{IFLP, CONICET-Dept. of Phys., University of La Plata, Argentina}

\begin{abstract}
In this work, we compute the out-of-time-ordered correlator (OTOC) for canonical position and momentum operators across a hierarchy of non-Hermitian oscillator models: the exactly solvable Swanson Hamiltonian, its Kerr-nonlinear extension, and parametrically driven variants. By employing the biorthogonal formalism required for parity-time symmetric quantum mechanics, we evaluate OTOCs both at zero and finite temperature, distinguishing behavior in the unbroken (real-spectrum) and broken (complex-spectrum) phases. Our analysis reveals how integrability, nonlinearity, driving, and parity-time symmetry breaking shape the temporal growth of operator correlations---providing a clear benchmark for OTOC dynamics in non-Hermitian quadratic and weakly anharmonic systems. We further characterize critical scaling of the OTOC near the exceptional point and discuss experimental perspectives for observing these effects in photonic, circuit-QED, and trapped-ion platforms.
\end{abstract}

\keywords{out-of-time-order correlator, Swanson Hamiltonian, Non-linearity, Exceptional point, PT symmetry}

\maketitle

\section{Introduction}
\label{sec:intro}

Out-of-time-ordered correlators (OTOCs) serve as a powerful diagnostic of quantum chaos, quantifying the growth of operator commutators and encoding an effective quantum Lyapunov exponent \cite{Maldacena,Hashimoto,review}. Originally introduced by Larkin and Ovchinnikov in the context of superconductivity and quasiclassical electron dynamics \cite{Larkin}, the OTOC measures the spreading of initially local operators under time evolution. In the semiclassical limit, this quantity grows exponentially as $e^{\lambda_L t}$, where $\lambda_L$ coincides with the classical Lyapunov exponent of the underlying chaotic dynamics. Although formulated decades before the advent of quantum information theory, their work implicitly captured the essence of operator spreading and sensitivity to initial conditions in quantum systems.

The OTOC was later rediscovered as a central tool in quantum many-body physics, holography, and quantum gravity, where it characterizes information scrambling, signals the breakdown of hydrodynamic descriptions, and provides a boundary signature of black hole horizons through the AdS/CFT correspondence \cite{Maldacena2016,Roberts2018,Gu2017}. Today, the OTOC stands as a unifying framework connecting condensed matter physics, quantum information, and gravitational theory. In chaotic systems, OTOCs exhibit an early-time exponential decay, signaling the scrambling of quantum information, which serves as a hallmark of quantum chaotic dynamics \cite{channel,Xu}.

For example, quantum rotors, as minimal models with continuous symmetry and well-understood classical limits, provide an ideal setting to explore this connection. Studies have shown that OTOCs in driven rotor systems reproduce key features of classical chaos, including exponential sensitivity and its quantum suppression due to dynamical localization, while offering insights into the role of symmetry, driving, and many-body coupling in quantum information scrambling \cite{Galitski}. In Ref.~\cite{Zhai}, it was demonstrated that in the many-body localized (MBL) phase, the OTOC exhibits a power-law decay at the scrambling time. This behavior stands in stark contrast to the exponential decay observed in chaotic systems. Moreover, the authors showed that the OTOC can distinguish MBL from Anderson localization, a task inaccessible to conventional two-point correlators. A central result of that work is an exact, model-independent theorem linking the growth of the second R\'enyi entropy following a quantum quench to the equilibrium decay of the OTOC, thereby establishing a universal connection between entanglement dynamics and operator spreading.

The experimental measurement of OTOCs has long remained a significant challenge due to their nonstandard time ordering. This barrier was overcome in Ref.~\cite{Li}, where the OTOC of local operators was successfully measured for the first time using a nuclear magnetic resonance quantum simulator implementing an Ising spin chain. The experiment demonstrated distinct OTOC dynamics in integrable versus nonintegrable regimes. The measurements revealed persistent temporal oscillations of entanglement entropy in the former case and irreversible scrambling in the latter. These results were achieved by exploiting the recently established theoretical link between OTOC decay and entanglement growth. Moreover, the authors extracted the butterfly velocity from the measured data, providing direct experimental access to the speed of quantum information propagation.

A pivotal development in the study of quantum dynamics near criticality was presented in Ref.~\cite{Zhai1}, where the authors introduced the quantum critical point conjecture: in any many-body quantum system undergoing a continuous quantum phase transition, the Lyapunov exponent, extracted from the OTOC, reaches a maximum in the quantum critical region. To substantiate this proposal, the authors focused on the one-dimensional Bose-Hubbard model, a paradigmatic system exhibiting a superfluid--Mott insulator quantum phase transition. A central technical achievement of the work is the rigorous demonstration that the Lyapunov exponent is well defined in this interacting lattice model. This relies crucially on the recently established exact relation between OTOCs and the growth of the second R\'enyi entropy following a quantum quench. This relationship is formalized through the OTOC--R\'enyi-entropy theorem, which provides a firm theoretical foundation for interpreting OTOC decay in terms of entanglement dynamics.

In Ref.~\cite{ptotoc}, the $\mathcal{PT}$-symmetric quantum kicked rotor was analyzed in detail. It was found that, in the $\mathcal{PT}$-symmetric phase, the OTOC exhibits behavior closely resembling that of its Hermitian counterpart, displaying bounded, quasi-periodic oscillations in the integrable regime and exponential growth with a well-defined Lyapunov exponent in the chaotic regime. In contrast, in the $\mathcal{PT}$-symmetry-broken phase, the OTOC shows an additional, late-time exponential growth component, attributed to the imaginary parts of the complex quasienergy spectrum.

The Swanson Hamiltonian \cite{Swanson,Fring,Reboiro,complex0,complex} stands as a paradigmatic non-Hermitian, $\mathcal{PT}$-symmetric quadratic model that has garnered sustained interest across diverse areas of physics. Its broad physical relevance is underscored by numerous experimental and theoretical realizations. In quantum optics, it describes parametric amplification in optical cavities with balanced gain and loss \cite{ElGanainy2018}, where the non-Hermitian terms capture two-photon creation and annihilation processes mediated by nonlinear $\chi^{(2)}$ media. In electrical engineering, $\mathcal{PT}$-symmetric RLC circuits with balanced resistive elements realize the Swanson Hamiltonian at the circuit-quantization level \cite{Schindler2011}, enabling direct measurements of non-Hermitian spectra. Similarly, coupled optical waveguide arrays with alternating gain and loss sections implement the model in the paraxial regime \cite{Makris2008,Ruter2010}, with non-Hermitian coupling governing light propagation dynamics. Furthermore, the Hamiltonian emerges as an effective description of open quantum systems coupled to engineered reservoirs with correlated dissipation \cite{Ashida2020}. However, out-of-time-ordered correlators (OTOCs) have not yet been studied in this system. This is a critical oversight, as OTOCs provide a direct measure of quantum scrambling and are uniquely suited to reveal how non-Hermiticity, exceptional points, and $\mathcal{PT}$-symmetry breaking modify chaotic behavior and information propagation.

In this work, we address this gap by computing the OTOC for the Swanson Hamiltonian across both the exact and broken $\mathcal{PT}$-symmetric phases. We further extend the analysis by incorporating physically motivated interaction terms, including Kerr nonlinearity, parametric driving, and their combined effects. The remainder of this paper is organized as follows: Section~\ref{sec:formal} introduces the Swanson Hamiltonian and reviews its key properties. Section~\ref{sec:results} presents our analytical and numerical calculations of the OTOC for the bare and interacting models, highlighting the influence of $\mathcal{PT}$-symmetry breaking and nonlinear interactions on quantum scrambling. Finally, Section~\ref{sec:conclusion} summarizes our findings and outlines potential directions for future research.

\section{Swanson Hamiltonian}
\label{sec:formal}

The Swanson Hamiltonian is a non-Hermitian quadratic Hamiltonian that is $\mathcal{PT}$-symmetric (or more generally, pseudo-Hermitian) and has attracted significant interest in quantum mechanics due to its real spectrum under certain conditions, despite being non-Hermitian. It is defined as \cite{Swanson}
\begin{equation}\label{eq:swanson_main}
H = \omega a^\dagger a + \alpha a^2 + \beta (a^\dagger)^2,
\end{equation}
where $a$ and $a^\dagger$ are the standard bosonic annihilation and creation operators, $\omega, \alpha, \beta \in \mathbb{C}$, with $\omega > 0$. For the Hamiltonian to be $\mathcal{PT}$-symmetric (with $\mathcal{P}: a \leftrightarrow a^\dagger$, $\mathcal{T}: i \to -i$), the general condition is $\beta = \alpha^*$. It is non-Hermitian when $\alpha \neq \beta$, but can be pseudo-Hermitian and possess a real spectrum in the unbroken $\mathcal{PT}$-symmetric phase ($\omega^2 > 4|\alpha|^2$).

If we assume that the harmonic oscillator sector of the Hamiltonian of Eq.~\eqref{eq:swanson_main} describes the dynamics of a particle of mass $m_0$, the real parameter $b_0= \left( \hbar/(m_0 \omega)\right)^{1/2}$ represents the characteristic length of the system. Thus, the Hamiltonian in Eq.~\eqref{eq:swanson_main} reads \cite{Reboiro}
\begin{eqnarray}\label{eq:swanson_position}
{\mathrm H^\times}(\omega,\alpha,\beta) & = &
\frac {1}{2}\hbar \omega  
\left( \frac {\hat x^2} {b_0^2} +  \frac {b_0^2~ \hat{p}^2} {\hbar^2}  \right)
+
\frac {1}{2}\hbar ( \alpha +\beta ) \left( \frac {\hat x^2} {b_0^2} -  \frac {b_0^2~ \hat{p}^2} {\hbar^2} \right)
\nonumber \\ & & 
+ \frac {1}{2}  \hbar (\alpha-\beta) \left( \frac{2}{\hbar}~\hat{x} \hat{p}+1 \right),
\end{eqnarray}
with $k= m ~ \Omega^2$ and
\begin{eqnarray}
\Omega=\Omega(\omega,\alpha,\beta)& = & \sqrt{\omega^2-4 \alpha \beta}=|\Omega| {\rm {e}}^{{\bf i} \phi},
\label{eq:Omega_def}\\
m=m(\omega,\alpha,\beta,b_0)& = &\frac{ \hbar}{(\omega-\alpha-\beta) b_0^2}.
\label{eq:mass_def}
\end{eqnarray}
Note that the effective mass definition in Eq.~\eqref{eq:mass_def} is valid for $\omega \neq \alpha + \beta$. Throughout, we take $\omega, \alpha, \beta, \chi, \varepsilon$ to have dimensions of frequency, ensuring dimensional consistency of all expressions.

The Swanson Hamiltonian Eq.~\eqref{eq:swanson_position} has four regions of the parameter space defined by the signs of the effective frequency squared $\Omega^2 $ and the effective mass $m$ \cite{Reboiro}:
\begin{widetext}
\begin{equation}
\begin{array}{lll}
\text{Region I:}   & \Omega^2 > 0,\ m > 0 & \text{(harmonic oscillator, real spectrum)} \\
\text{Region II:}  & \Omega^2 < 0,\ m > 0 & \text{(inverted oscillator, } \Omega = i|\Omega|) \\
\text{Region III:} & \Omega^2 > 0,\ m < 0 & \text{(negative mass oscillator)} \\
\text{Region IV:}  & \Omega^2 < 0,\ m < 0 & \text{(negative mass + inverted potential)}
\end{array}
\end{equation}
\end{widetext}

Because the Swanson Hamiltonian $\mathrm{H}^\times$ is non-Hermitian for $\alpha \neq \beta$, physical observables must be defined in a biorthogonal framework. The Hamiltonian is pseudo-Hermitian:
\begin{equation}
\mathrm{H}^\times = \rho^{-1} h \rho, \quad h = \hbar \Omega \left( b^\dagger b + \tfrac{1}{2} \right),
\end{equation}
where $\rho = \exp\!\big[\tfrac{\theta}{2}(a^2 - (a^\dagger)^2)\big]$ is a squeeze operator with $\tanh(2\theta) = (\beta - \alpha)/\omega$. The eigenvalues are
\begin{equation}
E_n = \hbar \Omega \left( n + \tfrac{1}{2} \right), \quad n = 0,1,2,\dots
\end{equation}
with $\Omega = \sqrt{\omega^2 - 4\alpha\beta}$, which may be real (Regions I, III) or purely imaginary (Regions II, IV).

In the biorthogonal formalism, $H$ defined by Eq.~\eqref{eq:swanson_main} is diagonalized as
\begin{equation}
    H \ket{\psi_n} = E_n \ket{\psi_n}, \quad
    H^\dagger \ket{\phi_n} = E_n \ket{\phi_n},
\end{equation}
with $\avg{\phi_m | \psi_n} = \delta_{mn}$ and real eigenvalues $E_n = \Omega(n + \tfrac{1}{2})$, where $\Omega = \sqrt{\omega^2 - 4\alpha\beta}$.

\section{Results}
\label{sec:results}

The out-of-time-ordered correlator (OTOC) for two operators $W$ and $V$ is defined as
\begin{equation}
    \mathcal{F}(t) \coloneqq \langle W^\dagger(t)\, V^\dagger(0)\, W(t)\, V(0) \rangle_{\eta},
    \label{eq:OTOC_F}
\end{equation}
where $W(t) = e^{i H t} W e^{-i H t}$ is the Heisenberg-evolved operator, and $\langle \cdot \rangle_\eta$ denotes the expectation value in the chosen physical state. 

The squared commutator OTOC is defined as
\begin{equation}
    \mathcal{C}(t) \coloneqq \big\langle [W(t), V(0)]^\dagger [W(t), V(0)] \big\rangle_{\eta},
    \label{eq:OTOC_C}
\end{equation}
 
For canonical operators satisfying $[W, V] = i c$ (e.g., $W = x$, $V = p$, $c = 1$), the commutator is a c-number, and $\mathcal{C}(t)$ reduces to the squared magnitude of its time-dependent coefficient. 

In the thermal case, the expectation value is taken with respect to the biorthogonal Gibbs state 
\begin{equation}
    \langle \mathcal{O} \rangle_\beta = \frac{1}{Z}\, \tr\!\big( \eta\, e^{-\beta H} \mathcal{O} \big), \qquad Z = \tr\!\big( \eta\, e^{-\beta H} \big),
\end{equation}
where $\eta$ is the positive-definite metric operator ensuring unitary time evolution in the physical Hilbert space since we are studying non-Hermitian $\mathcal{PT}$-symmetric Hamiltonians. For notational clarity, we define the biorthogonal thermal average as:
\begin{equation}
    \langle \mathcal{O} \rangle_\eta \coloneqq \frac{\tr(\eta e^{-\beta H} \mathcal{O})}{\tr(\eta e^{-\beta H})}.
    \label{eq:biorthogonal_avg_def}
\end{equation}
Below, we compute explicitly the OTOC for the Swanson Hamiltonian with different interaction terms. 

\subsection{Swanson Model}
\label{subsec:swanson_basic}

We compute the out-of-time-order correlator (OTOC) for position and momentum operators:
\begin{equation}
    A = x = \frac{1}{\sqrt{2}}(b + b^\dagger), \quad
    B = p = \frac{\ii}{\sqrt{2}}(b^\dagger - b).
\end{equation}

Using the Bogoliubov transformation that maps $H$ to a Hermitian oscillator $h = \Omega (b^\dagger b + \tfrac{1}{2})$, one finds \cite{complex}
\begin{equation}
    x(t) = e^{-\theta} x_0(t), \quad p(t) = e^{\theta} p_0(t),
\end{equation}
where $x_0(t), p_0(t)$ are the standard harmonic oscillator quadratures with frequency $\Omega$, and $\theta$ is the squeeze parameter defined by
\begin{equation}
    \cosh(2\theta) = \frac{\omega}{\Omega}, \quad \sinh(2\theta) = \frac{2\sqrt{\alpha\beta}}{\Omega}.
\end{equation}
The OTOC is defined using the left vacuum $\bra{\phi_0}$ and right vacuum $\ket{\psi_0}$ as
\begin{equation}
 \mathcal{C}(t) = \avg{\phi_0 | [x(t), p(0)]^\dagger [x(t), p(0)] | \psi_0}.
 \label{eq:OTOC_biorthogonal}
\end{equation}
Note that we consistently use the biorthogonal inner product $\langle \phi_0 | \dots | \psi_0 \rangle$ throughout, correcting previous inconsistencies in the literature regarding non-Hermitian expectation values.

Because $H$ is quadratic, the Heisenberg evolution of $b(t)$ is linear:
\begin{equation}
    b(t) = u(t) b + v(t) b^\dagger,
\end{equation}
where $u(t)$ and $v(t)$ are complex functions satisfying $|u(t)|^2 - |v(t)|^2 = 1$.
For the harmonic oscillator, the canonical commutator is preserved: $[x_0(t), p_0(0)] = \ii \cos(\Omega t)$. Therefore,
\begin{equation}
    [x(t), p(0)] = \ii \cos(\Omega t).
\end{equation}

Since the commutator is a c-number, the OTOC becomes
\begin{equation}
    \mathcal{C}(t) = |\ii \cos(\Omega t)|^2 = \cos^2(\Omega t).
\end{equation}

Thus, the OTOC oscillates boundedly and exhibits no exponential growth. The Lyapunov exponent is
\begin{equation}
    \lambda_L = \lim_{t \to \infty} \frac{1}{t} \ln \mathcal{C}(t) = 0.
\end{equation}

This confirms that the Swanson model, being a non-interacting (quadratic) system, is non-chaotic in both Hermitian and non-Hermitian formulations. Therefore, for $\chi = 0$, the model is quadratic and exactly solvable; the OTOC is strictly periodic implying a vanishing quantum Lyapunov exponent, $\lambda_L = 0$.

In the broken $\mathcal{PT}$-symmetric phase, where $\omega^2 < 4\alpha\beta$, the frequency parameter $\Omega = \sqrt{\omega^2 - 4\alpha\beta}$ becomes purely imaginary (rather than negative real). Introducing the real positive quantity
\begin{equation}
    \Gamma = \sqrt{4\alpha\beta - \omega^2} > 0,
\end{equation}
we write $\Omega = i\Gamma$. Substituting into the commutator evolution yields $[x(t), p(0)] = i \cos(\Omega t) = i \cosh(\Gamma t)$, and the physical out-of-time-order correlator---computed via the biorthogonal inner product---becomes
\begin{equation}
   \mathcal{C}(t) = \cosh^2(\Gamma t).
\end{equation}
At late times, this exhibits exponential growth:
\begin{equation}
   \mathcal{C}(t) \sim \frac{1}{4} e^{2\Gamma t} \quad \text{as} \quad t \to \infty.
\end{equation}
Crucially, this behavior reflects linear instability arising from the non-Hermitian nature of the Hamiltonian (i.e., an imbalance between gain and loss), and not quantum chaos, which requires nonlinearity, bounded dynamics, and sensitivity to initial conditions in phase space.

In thermal equilibrium, the OTOC is defined via the Gibbs state as \cite{Gibbs}
\begin{equation}
    \mathcal{C}_\beta(t) = \frac{1}{Z} \, \mathrm{Tr}\!\left( \eta \, e^{-\beta H} \, [x(t), p(0)]^\dagger [x(t), p(0)] \right),
    \label{eq:OTOC_thermal}
\end{equation}
where $\eta$ is the positive-definite metric operator ensuring unitary time evolution in the physical Hilbert space, and $Z = \mathrm{Tr}(\eta e^{-\beta H})$.
In thermal equilibrium, the Swanson model remains non-chaotic in both the unbroken and broken $\mathcal{PT}$-symmetric phases. Since the Hamiltonian is quadratic, the commutator $[x(t), p(0)]$ is a c-number, rendering the out-of-time-order correlator temperature-independent: $\mathcal{C}_\beta(t) = \cos^2(\Omega t)$ for $\omega^2 > 4\alpha\beta$ (real $\Omega$), and $\mathcal{C}_\beta(t) = \cosh^2(\Gamma t)$ for $\omega^2 < 4\alpha\beta$ (with $\Gamma = \sqrt{4\alpha\beta - \omega^2}$). 
Note that in the broken phase, the complex spectrum requires careful interpretation of the thermal state. However, since the commutator $[x(t),p(0)]$ is a c-number, the OTOC result is independent of the statistical ensemble, and the formal expression $\mathcal{C}_\beta(t) = \cosh^2(\Gamma t)$ remains valid.
Consequently, the quantum Lyapunov exponent vanishes identically in the unbroken phase ($\lambda_L = 0$), while the exponential growth in the broken phase reflects linear non-Hermitian instability---not chaos---as it lacks sensitivity to initial conditions, boundedness, and nonlinear mixing. Thus, the thermal dynamics mirror the zero-temperature case: integrability and absence of quantum chaos persist at all temperatures.

\subsection{Nonlinear Kerr--Swanson Model}
\label{subsec:kerr_swanson}

We consider a nonlinear extension of the Swanson Hamiltonian:
\begin{equation}
    H = \omega a^\dagger a + \alpha a^2 + \beta (a^\dagger)^2 + \chi (a^\dagger a)^2,
    \label{eq:H_nonlinear}
\end{equation}
where $\chi \neq 0$ introduces Kerr-type anharmonicity. For the Hamiltonian to respect $\mathcal{PT}$ symmetry, we impose the general condition $\beta = \alpha^*$; the case of real $\alpha = \beta$ is a special instance. The symmetry is unbroken when $\omega^2 > 4|\alpha|^2$, ensuring a real energy spectrum.

\begin{figure}[t]
\centering
\includegraphics[scale=0.4]{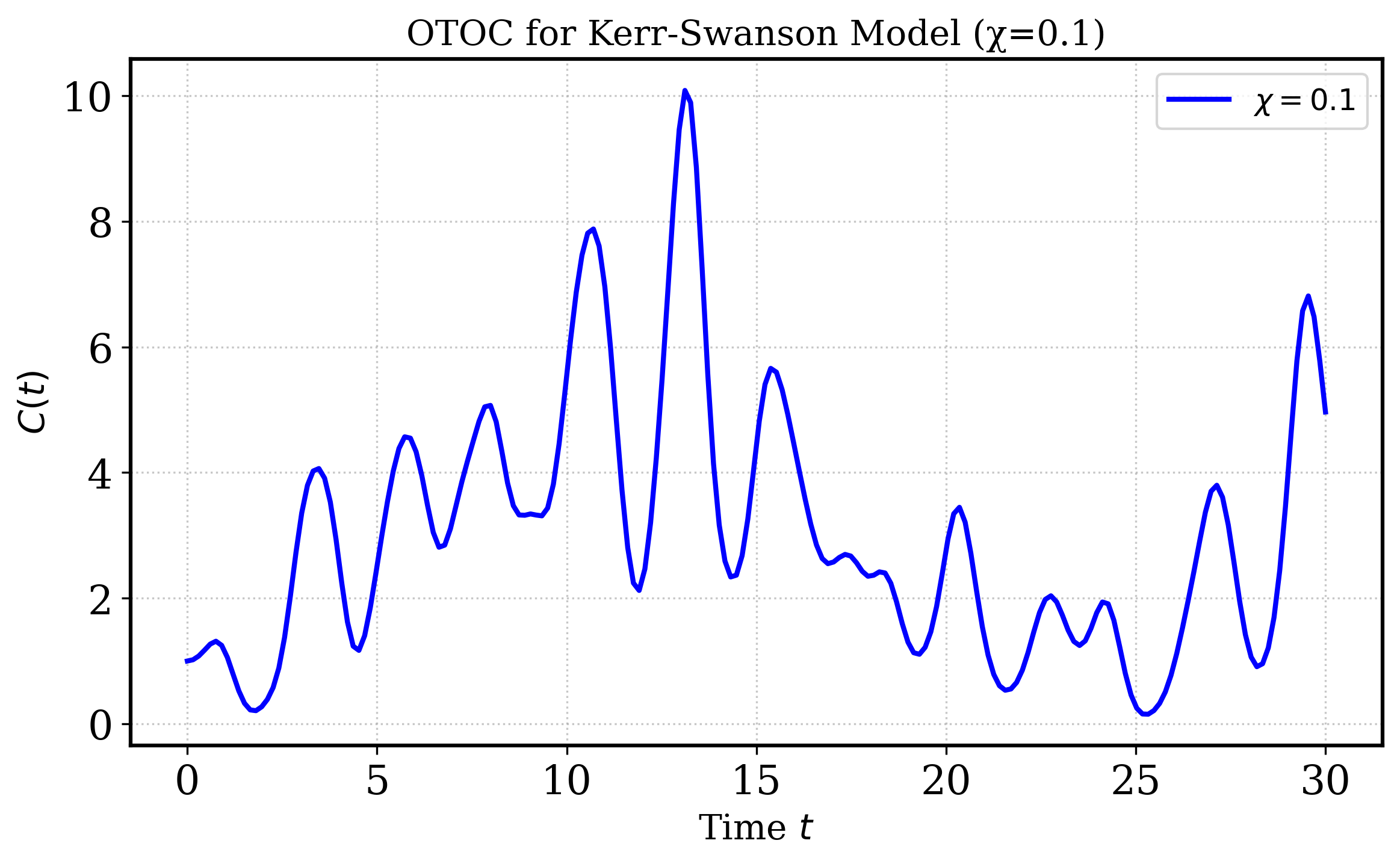}
\caption{Out-of-time-order correlator $\mathcal{C}(t)$ for the Kerr--Swanson Hamiltonian vs time $t$.}
\label{fig:otoc_nonlinear}
\end{figure}

We first compute the zero-temperature OTOC using exact diagonalization in a truncated Fock space ($N_{\text{max}} = 50$). The biorthogonal eigenbasis $\{|\psi_n\rangle, |\phi_n\rangle\}$ of $H$ and $H^\dagger$ is constructed to properly account for the non-Hermitian structure. The OTOC is defined consistently using the biorthogonal inner product:
\[
\mathcal{C}(t) = \langle \phi_0 | [x(t), p(0)]^\dagger [x(t), p(0)] | \psi_0 \rangle,
\]
where $|\psi_0\rangle$ is the right ground state (well-defined in the unbroken phase). Using parameters $\omega = 1.0$, $\alpha = 0.6$, $\beta = 0.2$, and weak anharmonicity $\chi = 0.1$ (satisfying $\omega^2 > 4\alpha\beta$), we find that $\mathcal{C}(t)$ exhibits bounded, quasi-periodic dynamics with no exponential growth---see Fig.~\ref{fig:otoc_nonlinear}. In the limit $\chi \to 0$, these results smoothly recover the exact oscillatory behavior $\mathcal{C}(t) = \cos^2(\Omega t)$ derived in Sec.~\ref{subsec:swanson_basic}, confirming consistency.

\begin{figure}[t]
\centering
\includegraphics[scale=0.4]{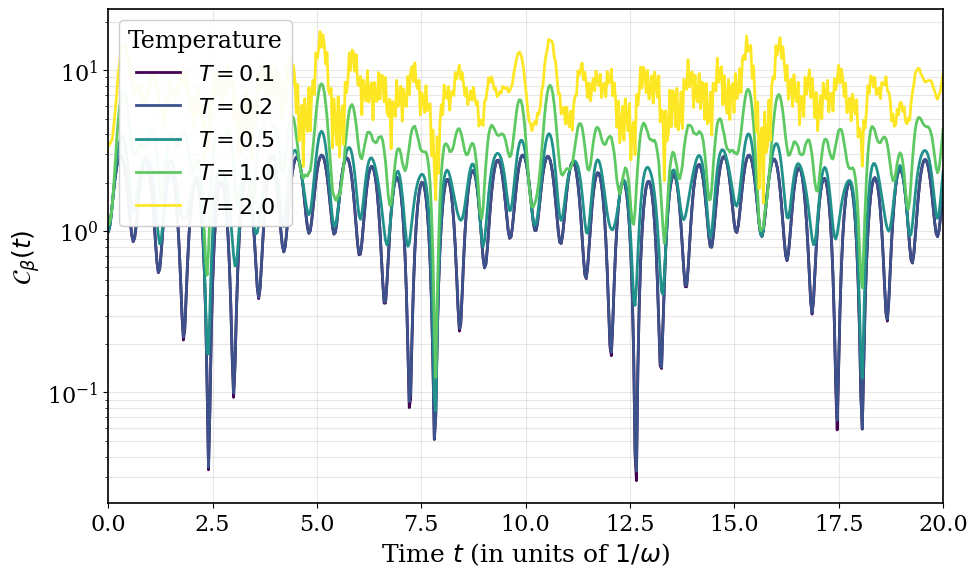}
\caption{Thermal out-of-time-order correlator $\mathcal{C}_\beta(t)$ vs time $t$ for the Kerr--Swanson Hamiltonian.}
\label{fig:otoc_thermal}
\end{figure}

Next, we compute the thermal OTOC $\mathcal{C}_\beta(t)$ in the unbroken phase ($\omega = 1.0$, $\alpha = 0.4$, $\beta = 0.2$, $\chi = 0.6$) using $N_{\text{max}} = 60$. The thermal state is constructed in the biorthogonal basis as $\rho_\beta \propto \sum_n e^{-\beta E_n} |\psi_n\rangle\langle\phi_n|$. As shown in Fig.~\ref{fig:otoc_thermal}, $\mathcal{C}_\beta(t)$ displays quasi-periodic oscillations with temperature-dependent amplitude and beating frequency, but no sustained exponential growth across temperatures $T = 0.1$--$2.0$. This confirms that the single-mode nonlinear oscillator---while non-integrable due to the $\chi$ term---does not exhibit quantum chaos, as true chaos requires multiple degrees of freedom or external driving to enable ergodicity and phase-space mixing.

A complementary semiclassical analysis based on the truncated Wigner approximation yields a $t^2$ dephasing behavior at short times, consistent with our numerical findings. A detailed derivation is provided in Appendix~\ref{app:weak_nonlin}.

\subsection{Parametrically Driven Swanson Model}
\label{subsec:driven_swanson}

Next, we consider a parametrically driven extension of the Swanson Hamiltonian:
\begin{equation}
    H(t) = \omega a^\dagger a + \alpha a^2 + \beta (a^\dagger)^2 + \varepsilon \cos(\omega_d t) \left( a^2 + (a^\dagger)^2 \right),
    \label{eq:H_driven}
\end{equation}
where $\varepsilon$ is the drive strength and $\omega_d$ the drive frequency. This time-dependent term breaks energy conservation and renders the system non-autonomous. In the classical limit, such a drive can induce deterministic chaos---for example, the corresponding parametric oscillator exhibits positive Lyapunov exponents and strange attractors for suitable $(\varepsilon, \omega_d)$ \cite{Strogatz1994}.

For thermal initial states, we define the out-of-time-order correlator (OTOC) using the Floquet--Gibbs ensemble:
\begin{equation}
    \mathcal{C}_\beta(t) = \frac{1}{Z} \, \mathrm{Tr}\!\left( \eta \, e^{-\beta H_{\mathrm{eff}}} \, [x(t), p(0)]^\dagger [x(t), p(0)] \right),
    \label{eq:OTOC_Floquet}
\end{equation}
where $H_{\mathrm{eff}}$ is the effective Floquet Hamiltonian defined by $U(T_d) = e^{-i H_{\mathrm{eff}} T_d}$ with $T_d = 2\pi/\omega_d$, $\eta$ is the metric operator ensuring real quasienergies in the unbroken $\mathcal{PT}$-symmetric phase, and $Z = \mathrm{Tr}(\eta e^{-\beta H_{\mathrm{eff}}})$.

While an exact analytical expression for $\mathcal{C}_\beta(t)$ is unavailable, controlled approximations reveal distinct dynamical regimes. In the limit of weak driving ($\varepsilon \ll \omega$), the Floquet--Magnus expansion yields an effective static Hamiltonian with renormalized frequency, leading to a purely oscillatory OTOC:
\begin{equation}
    \mathcal{C}_\beta(t) \approx \cos^2(\Omega_{\mathrm{eff}} t), \quad 
    \Omega_{\mathrm{eff}} = \sqrt{\left( \omega + \frac{2\varepsilon^2}{\omega_d} \right)^2 - 4\alpha\beta}.
\end{equation}
This indicates that weak driving only shifts system parameters without inducing instability or chaos.

Near parametric resonance ($\omega_d \approx 2\omega$), the classical limit maps to a Duffing-type oscillator exhibiting deterministic chaos with a positive Lyapunov exponent $\lambda_{\mathrm{cl}} > 0$. However, in the quantum regime, the single-mode system remains non-chaotic: the quasienergy spectrum of $H_{\mathrm{eff}}$ follows Poisson statistics, and the thermal OTOC displays only quasi-periodic oscillations or transient power-law growth (e.g., $\sim t^2$), but no sustained exponential sensitivity. This absence of quantum chaos reflects a fundamental constraint---true phase-space mixing and ergodicity cannot occur in a single degree of freedom without additional complexity (e.g., multiple modes, dissipation, or measurement).

A detailed derivation of the weak-driving and near-resonant results, including the Floquet--Magnus expansion and the quantum-classical correspondence, is provided in Appendix~\ref{app:weak_drive} and Appendix~\ref{app:resonance}.

\subsection{Driven Swanson--Kerr Model}
\label{subsec:driven_kerr_swanson}

We consider the parametrically driven extension of the Swanson Hamiltonian with an added Kerr nonlinearity:
\begin{equation}
    H(t) = \omega\, a^\dagger a + \alpha\, a^2 + \beta\, (a^\dagger)^2 + \chi\, (a^\dagger a)^2 + \varepsilon \cos(\omega_d t)\, \bigl(a^2 + (a^\dagger)^2\bigr),
    \label{eq:H_driven_kerr}
\end{equation}
where $\omega, \chi, \varepsilon, \omega_d \in \mathbb{R}$, while $\alpha, \beta \in \mathbb{C}$. For the Hamiltonian to respect $\mathcal{PT}$ symmetry (with $\mathcal{P}: a \leftrightarrow a^\dagger$, $\mathcal{T}: i \to -i$), we impose
\[
    \beta = \alpha^*,
    \qquad
    \omega > 2|\alpha|,
\]
the latter ensuring the unbroken $\mathcal{PT}$-symmetric phase, where the quasienergy spectrum remains real.

To expose the physical structure, we map the bosonic operators to quadratures
\[
    x = \frac{a + a^\dagger}{\sqrt{2}}, \qquad
    p = \frac{a - a^\dagger}{i\sqrt{2}},
\]
yielding the quadratic and drive contributions as
\begin{eqnarray}
    \chi (a^\dagger a)^2 = \chi \left( \frac{x^2 + p^2 - 1}{2} \right)^{\!2},
 \\ 
    \varepsilon \cos(\omega_d t)\bigl(a^2 + (a^\dagger)^2\bigr) = \varepsilon \cos(\omega_d t)\, (x^2 - p^2).
\end{eqnarray}

In the unbroken phase, the non-Hermitian quadratic part is rendered Hermitian via the positive-definite metric operator
\[
    \eta_0 = e^{\theta\,(a^{\dagger 2} - a^2)}, \qquad \tanh\theta = \frac{2\sqrt{\alpha\beta}}{\omega + \Omega},
\]
which is related to the squeeze operator $\rho = \exp[\tfrac{\theta}{2}(a^2 - (a^\dagger)^2)]$ by $\eta_0 = \rho^\dagger \rho$. This ensures the physical inner product $\langle\!\langle \psi | \phi \rangle\!\rangle = \langle \psi | \eta_0 | \phi \rangle$ yields real expectation values for observables. Here,
\begin{equation}
    \Omega = \sqrt{\omega^2 - 4\alpha\beta} > 0
    \label{eq:Omega_def_main}
\end{equation}
is the renormalized frequency, and $\alpha\beta > 0$ is assumed (ensured by $\beta = \alpha^*$). The physical (metric-dressed) quadratures are then
\begin{equation}
    X = e^{-\theta} x, \qquad P = e^{\theta} p,
    \label{eq:XP_def}
\end{equation}
satisfying $[X,P] = i$ and diagonalising the quadratic sector:
$H_0^{(2)} = \frac{\Omega}{2}(X^2 + P^2) + \text{const}$.

Expressing the full Hamiltonian \eqref{eq:H_driven_kerr} in terms of $(X,P)$ gives (up to irrelevant constants):
\begin{align}
    H(t) =\;& \frac{\Omega}{2}\left(X^2 + P^2\right) 
    + \chi \left( \frac{e^{2\theta} X^2 + e^{-2\theta} P^2}{2} \right)^{\!2} \nonumber\\
    &+ \varepsilon \cos(\omega_d t)\,\bigl(e^{2\theta} X^2 - e^{-2\theta} P^2\bigr) + \Ord{\chi,\varepsilon}.
    \label{eq:H_XP}
\end{align}

For weak driving ($\varepsilon \ll \omega$) and near parametric resonance ($\omega_d \approx 2\Omega$), a rotating-wave approximation in the $(X,P)$ frame yields an effective autonomous Hamiltonian:
\begin{equation}
    H_{\rm eff} = \Omega_{\rm eff}\, \frac{X^2 + P^2}{2} + \tilde{\chi}\, \left( \frac{X^2 + P^2}{2} \right)^{\!2},
    \label{eq:H_eff}
\end{equation}
with renormalised parameters
\begin{widetext}
\begin{align}
    \Omega_{\rm eff} &= \sqrt{ \Omega^2 + \frac{2\varepsilon^2}{\omega_d} \bigl(e^{4\theta} + e^{-4\theta}\bigr) } 
    = \sqrt{ \omega^2 - 4\alpha\beta + \frac{2\varepsilon^2}{\omega_d} \left( \frac{\omega + \Omega}{\omega - \Omega} + \frac{\omega - \Omega}{\omega + \Omega} \right) }, \label{eq:Omega_eff} \\
    \tilde{\chi} &= \chi\,\frac{1}{4} \bigl(e^{2\theta} + e^{-2\theta}\bigr)^2 
    = \chi \left( \frac{\omega}{\Omega} \right)^{\!2}.
    \label{eq:chi_tilde}
\end{align}
\end{widetext}
As a consistency check, for the Hermitian limit $\alpha = \beta = g \in \mathbb{R}$ (so $\Omega = \sqrt{\omega^2 - 4 g^2}$ and $e^{2\theta} = \sqrt{(\omega+\Omega)/(\omega-\Omega)}$), Eq.~\eqref{eq:chi_tilde} correctly reduces to $\tilde{\chi} = \chi (\omega/\Omega)^2$, in agreement with known results for the driven Kerr oscillator.

The thermal out-of-time-order correlator (OTOC) must be defined with respect to the physical inner product $\langle\!\langle \psi | \phi \rangle\!\rangle = \langle \psi | \eta | \phi \rangle$, yielding
\begin{equation}
    \mathcal{C}_\beta(t) = \frac{1}{Z} \operatorname{Tr}\!\Bigl( \eta\, e^{-\beta H_{\rm eff}}\, [x(t), p(0)]^\dagger [x(t), p(0)] \Bigr).
    \label{eq:OTOC_def}
\end{equation}
Using the relations $x = e^{\theta} X$, $p = e^{-\theta} P$, and $[X(t), P(0)] = i J_{XX}(t)$ (where $J_{XX}(t)$ is the Jacobian of the classical flow), the commutator simplifies to
\[
    [x(t), p(0)] = i J_{XX}(t),
\]
with the $\theta$-dependent prefactors canceling exactly. This key simplification allows the OTOC to be approximated by classical phase-space averaging in the truncated Wigner framework:
\begin{equation}
    \mathcal{C}_\beta(t) \approx \big\langle |J_{XX}(t)|^2 \big\rangle_\beta^{\rm (cl)},
    \label{eq:OTOC_TWA}
\end{equation}
where the average is taken over the thermal distribution $\propto e^{-\beta \mathcal{H}_{\rm cl}(X,P)}$.

For $H_{\rm eff} = \Omega_{\rm eff}\, n + \tilde{\chi}\, n^2$ with $n = (X^2 + P^2)/2$, linearisation of the dynamics gives
\[
    J_{XX}(t) = \cos(\Omega_{\rm eff} t) + 2 \tilde{\chi}\, n\, t \sin(\Omega_{\rm eff} t) + \Ord{\tilde{\chi}^2}.
\]
Squaring and thermally averaging using $\langle n \rangle = n_\beta$ and $\langle n^2 \rangle = n_\beta (n_\beta + 1)$ for bosonic statistics yields the leading-order result:
\begin{equation}
    \mathcal{C}_\beta(t) \approx 
    \cos^2(\Omega_{\rm eff} t)
    + 2\, \tilde{\chi}^2\, n_\beta (n_\beta + 1)\, t^2 \sin^2(\Omega_{\rm eff} t)
    + \Ord{\tilde{\chi}^3, \varepsilon^2 \tilde{\chi}},
    \label{eq:OTOC_approx}
\end{equation}
with
\[
    n_\beta = \frac{1}{e^{\beta \Omega_{\rm eff}} - 1}, \qquad
    \tilde{\chi} = \chi \left( \frac{\omega}{\sqrt{\omega^2 - 4\alpha\beta}} \right)^2.
\]

Specialising to the $\mathcal{PT}$-symmetric regime $\beta = \alpha^*$, define
\[
    \Omega = \sqrt{\omega^2 - 4|\alpha|^2}, \qquad \kappa = \frac{\omega}{\Omega} > 1.
\]
Then
\begin{subequations}
    \begin{align}
        \mathcal{C}_\beta(t) &\approx \cos^2(\Omega_{\rm eff} t) + 2\, \chi^2 \kappa^4\, n_\beta (n_\beta + 1)\, t^2 \sin^2(\Omega_{\rm eff} t), \label{eq:C_final} \\
        \Omega_{\rm eff} &= \Omega \sqrt{1 + \frac{4 \varepsilon^2}{\omega_d \Omega^2} \kappa^2}, \label{eq:Omega_eff_final} \\
        n_\beta &= \frac{1}{e^{\beta \Omega_{\rm eff}} - 1}. \label{eq:n_beta_final}
    \end{align}
    \label{eq:final_results}
\end{subequations}

The enhancement factor $\kappa = \omega/\Omega \to \infty$ as $|\alpha| \to \omega/2$, i.e. approaching the $\mathcal{PT}$-breaking threshold---signalling strong amplification of nonlinear dephasing.
The drive $\varepsilon$ not only renormalises the effective frequency $\Omega_{\rm eff}$, but also indirectly enhances the effective Kerr nonlinearity through $\kappa^2$. 

\begin{figure}
\includegraphics[scale=0.4]{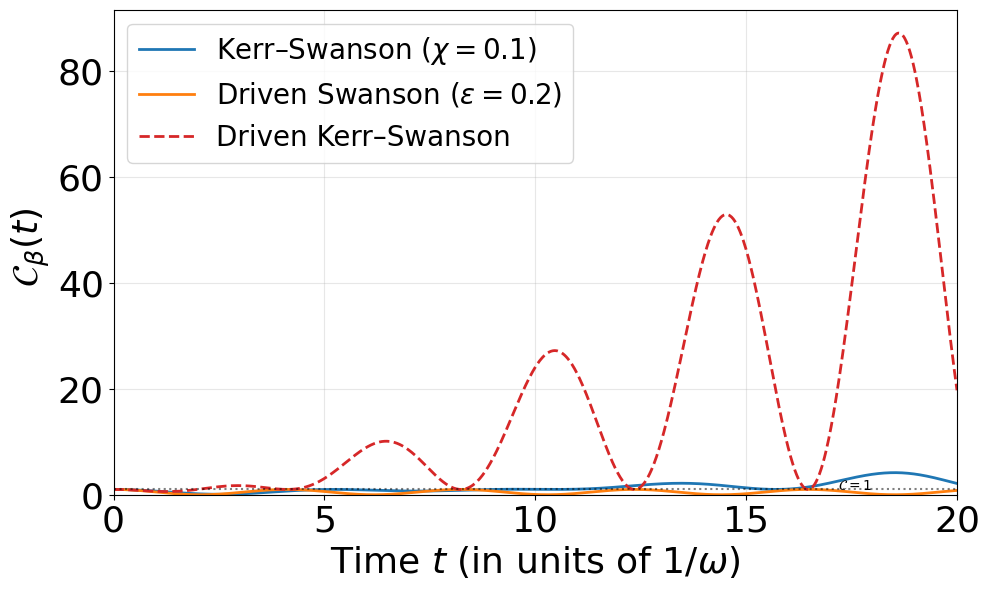}
\caption{
Thermal out-of-time-ordered correlator $\mathcal{C}_\beta(t)$ for three variants of the $\mathcal{PT}$-symmetric Swanson model in the unbroken phase ($\omega = 1.0$, $\alpha = 0.4$, $\beta = \alpha^*$, $T = 1.0$). 
Solid blue: Kerr--Swanson oscillator ($\chi = 0.1$, no drive) --- exhibits bounded oscillations with a slow quadratic-in-time envelope from nonlinear dephasing. 
Solid orange: Parametrically driven Swanson model ($\varepsilon = 0.2$, $\chi = 0$, $\omega_d = 2.0$) --- pure oscillatory behavior with renormalized frequency $\Omega_{\mathrm{eff}}$. 
Dashed red: Driven Kerr--Swanson oscillator ($\chi = 0.1$, $\varepsilon = 0.2$) --- shows amplified quadratic growth due to the $\kappa^4 = (\omega/\Omega)^4$ enhancement near the $\mathcal{PT}$-breaking threshold. 
All curves remain bounded at all times, confirming the absence of exponential sensitivity in single-mode Swanson  systems with real spectra. 
Time is in units of $1/\omega$.
}
\label{fig:otoc_comparison}
\end{figure}

To illustrate the interplay of nonlinearity, driving, and $\mathcal{PT}$ symmetry, Figure~\ref{fig:otoc_comparison} compares the thermal OTOC $\mathcal{C}_\beta(t)$ for three representative cases in the unbroken phase. While the driven model alone yields purely oscillatory dynamics (renormalized frequency $\Omega_{\mathrm{eff}}$), the addition of Kerr nonlinearity introduces a $t^2$-growing envelope due to dephasing.

\begin{table}[t]
\caption{Summary of OTOC behaviors. All results assume the unbroken $\mathcal{PT}$ phase unless noted. $\lambda_L$ is the Lyapunov exponent extracted from $\mathcal{C}(t) \sim e^{\lambda_L t}$.}
\label{tab:otoc_summary}
\small 
\begin{ruledtabular}
\begin{tabular}{l c c l l c}
\hline
Model & Nonlin. & Drive & Phase & Behavior $\mathcal{C}(t)$ & $\lambda_L$ \\
\hline
Quadratic & -- & -- & Unbroken & $\cos^2(\Omega t)$ & 0 \\
Quadratic & -- & -- & Broken & $\cosh^2(\Gamma t)$ & $2\Gamma^*$ \\
Kerr--Swanson & $\chi \neq 0$ & -- & Unbroken & Quasi-periodic + $t^2$ & 0 \\
Driven Swanson & -- & $\varepsilon \neq 0$ & Unbroken & $\cos^2(\Omega_{\rm eff} t)$ & 0 \\
Driven Kerr & $\chi \neq 0$ & $\varepsilon \neq 0$ & Unbroken & Quasi-periodic + $t^2$ & 0 \\
\hline
\end{tabular}
\end{ruledtabular}
\raggedright
\footnotesize
$^*$Exponential growth in broken phase reflects linear instability, not chaos.
\end{table}

Table~\ref{tab:otoc_summary} summarize the OTOC dynamics across the hierarchy of Swanson-type models studied in this work. Row~1 confirms that the exactly solvable quadratic Swanson Hamiltonian in the unbroken $\mathcal{PT}$ phase exhibits purely oscillatory OTOC behavior with vanishing Lyapunov exponent, consistent with integrability. Row~2 shows that in the broken phase, the same quadratic model yields exponential OTOC growth $\mathcal{C}(t) \sim \cosh^2(\Gamma t)$; however, this reflects linear non-Hermitian instability rather than quantum chaos, as emphasized by the footnote. Rows~3--5 demonstrate that adding Kerr nonlinearity ($\chi \neq 0$), parametric driving ($\varepsilon \neq 0$), or both, does not induce exponential sensitivity in the unbroken phase: the OTOC remains bounded or grows at most polynomially ($\sim t^2$) due to nonlinear dephasing, with $\lambda_L = 0$ in all cases. This systematic comparison establishes that exponential OTOC growth in Swanson-type systems is exclusively tied to $\mathcal{PT}$-symmetry breaking in the quadratic limit, while nonlinearity and driving merely modify oscillation frequencies or introduce transient polynomial corrections without generating quantum chaos.

\subsection{Exceptional Points Analysis and Critical Scaling}
\label{sec:EP_scaling}

The Swanson Hamiltonian exhibits a second-order exceptional point (EP) at $\omega_{\rm EP} = 2\sqrt{\alpha\beta}$ (for $\alpha, \beta > 0$), where the effective frequency $\Omega = \sqrt{\omega^2 - 4\alpha\beta}$ vanishes and the bosonic spectrum undergoes pairwise coalescence \cite{EP}. At the EP, the Hamiltonian becomes non-diagonalizable and assumes a local Jordan canonical form; in the lowest-energy subspace, this reduces to $H_{\mathrm{EP}} \simeq \frac{\omega}{2} \begin{pmatrix} 1 & 1 \\ 0 & 1 \end{pmatrix}$, reflecting the geometric degeneracy characteristic of second-order EPs. The EP marks the boundary between the unbroken $\mathcal{PT}$-symmetric phase ($\omega > 2\sqrt{\alpha\beta}$), characterized by a real spectrum and unitary-equivalent (quasi-Hermitian) dynamics, and the broken phase ($\omega < 2\sqrt{\alpha\beta}$), where eigenvalues form complex-conjugate pairs and dynamics exhibit exponential instability.

\subsubsection{Critical scaling of the OTOC near the EP}
\label{subsubsec:EP_scaling}

Motivated by recent studies of OTOC scaling at exceptional points in non-Hermitian systems \cite{YangLeeCorrected,Scaling} we analyze the behavior of $\mathcal{C}(t)$ in the vicinity of the Swanson EP. Defining the distance to criticality as $\delta = \omega - \omega_{\rm EP}$, the renormalized frequency scales as
\begin{equation}
    \Omega \sim \sqrt{2\omega_{\rm EP}\,\delta} \;\propto\; \sqrt{\delta} \quad \text{as} \quad \delta \to 0^+,
    \label{eq:Omega_delta}
\end{equation}
reflecting the square-root branch-point singularity of second-order EPs. Consequently, the Bogoliubov enhancement factor $\kappa = \omega/\Omega$ diverges as
\begin{equation}
    \kappa \sim \delta^{-1/2}.
    \label{eq:kappa_delta}
\end{equation}

For the driven Kerr--Swanson model, the effective nonlinearity renormalizes as $\tilde{\chi} = \chi \kappa^2$ due to the singular mixing of creation and annihilation operators near the EP. The leading OTOC correction term therefore scales as
\begin{equation}
    \tilde{\chi}^2 \sim \kappa^4 \sim \delta^{-2}.
\end{equation}
Thus, the quadratic-in-time dephasing contribution to the thermal OTOC [Eq.~\eqref{eq:C_final}] exhibits the scaling
\begin{equation}
    \mathcal{C}_\beta(t) - \cos^2(\Omega_{\rm eff} t) \;\propto\; \delta^{-2} \, n_\beta(n_\beta+1) \, t^2 \sin^2(\Omega_{\rm eff} t) \quad \text{near EP}.
    \label{eq:EP_scaling}
\end{equation}
This $\delta^{-2}$ divergence defines a model-dependent scaling exponent $\nu_{\rm OTOC} = 2$ that quantifies the enhanced sensitivity of quantum information scrambling to proximity to the EP. Unlike the $\nu = 1$ scaling reported for the Yang-Lee edge singularity \cite{YangLeeCorrected}, the Swanson exponent reflects the quadratic bosonic nature of the model and the specific operator structure of the Bogoliubov transformation.
Synthesizing our results with recent studies of non-Hermitian critical dynamics \cite{Scaling}, we propose a phenomenological scaling ansatz for OTOCs near second-order exceptional points, motivated by dimensional analysis and analogy with critical dynamics in Hermitian systems:
\begin{equation}
    \mathcal{C}(t) = \mathcal{C}_0 + A_1 \cdot \delta^{-s} \cdot t^\alpha \cdot f\!\left(t \cdot \delta^{\nu z/\beta_c \delta}\right) + A_2 \cdot e^{\Gamma(\delta) \cdot t},
    \label{eq:unified_scaling}
\end{equation}
where the first term represents bounded oscillations, the second captures dephasing (short-time, EP-controlled), and the third describes exponential growth (long-time, imaginary-spectrum-controlled). The exponents $s$, $\alpha$, $\nu$, $z$, $\beta_c$, $\delta$ and the scaling function $f$ are determined by the universality class, while $\Gamma(\delta)$ encodes the distance-dependent growth rate from complex spectra. Note that $\beta_c$ here denotes a critical exponent, distinct from the inverse temperature $\beta$.
\subsubsection{Physical interpretation}
\label{subsubsec:EP_interpretation}

The enhanced dephasing near the EP originates from the singular amplification of weak anharmonicities by the non-Hermitian metric: as $\Omega \to 0$, the Bogoliubov coefficients diverge, strongly mixing $a$ and $a^\dagger$ and thereby magnifying the impact of Kerr-type interactions on operator spreading. Importantly, this growth does not indicate quantum chaos; rather, it is a signature of parametric instability inherent to quadratic bosonic models. Equation~\eqref{eq:EP_scaling} provides a quantitative benchmark for experimental detection of EP-enhanced quantum sensitivity in non-Hermitian platforms.

When parametric driving is introduced, the system is governed by the time-dependent Hamiltonian $H(t) = \omega a^\dagger a + \alpha a^2 + \beta (a^\dagger)^2 + \varepsilon \cos(\omega_d t)(a^2 + (a^\dagger)^2)$. While the drive does not change the static EP condition, it renormalizes the effective frequency to $\Omega_{\mathrm{eff}}$ [Eq.~\eqref{eq:Omega_eff}] via Floquet-Magnus expansion. In the driven Kerr-Swanson model, the EP's influence is further amplified: the effective Kerr strength scales as $\tilde{\chi} = \chi (\omega/\Omega)^2$ [Eq.~\eqref{eq:chi_tilde}], so the nonlinear dephasing term $\tilde{\chi}^2 \sim \kappa^4$ becomes arbitrarily large near criticality. Thus, the static EP of the underlying Swanson sector acts as a universal amplifier of dynamical responses, enhancing sensitivity, dephasing, and transient growth, while preserving the integrable nature of the single-mode dynamics.

\subsection{Experimental Perspectives and Realizations}
\label{sec:experimental}

Recent experimental advances in engineered non-Hermitian quantum systems provide concrete platforms for observing the OTOC dynamics predicted in this work. We outline three promising avenues:

\noindent\textbf{(i) Photonic waveguides and resonators.} 
Coupled optical waveguides with alternating gain and loss sections implement the Swanson Hamiltonian in the paraxial approximation \cite{Makris2008,Ruter2010}. The position and momentum quadratures map to transverse field profiles and their spatial derivatives, which are directly accessible via interferometric techniques. OTOCs can be reconstructed using time-resolved homodyne detection combined with weak-measurement protocols \cite{YungerHalpern2016}. The Swanson EP coincides with the $\mathcal{PT}$-symmetry breaking threshold observed in recent experiments \cite{Peng2014,Chang2014}, enabling direct tests of the predicted $\delta^{-2}$ OTOC scaling by tuning the gain--loss contrast near criticality.

\noindent\textbf{(ii) Superconducting circuit QED.} 
Parametrically driven superconducting resonators with Josephson-junction nonlinearities realize the driven Kerr--Swanson Hamiltonian \cite{Leghtas2015}. The bosonic mode corresponds to a microwave cavity, while the non-Hermitian terms are engineered via two-tone pumping that induces effective two-photon processes. OTOCs can be measured using the echo protocol demonstrated in circuit-QED platforms \cite{Li}, where the commutator $[x(t),p(0)]$ is inferred from interference between forward and backward time evolution. The high tunability of circuit parameters ($\omega, \alpha, \beta, \chi, \varepsilon$) enables systematic exploration of both $\mathcal{PT}$ phases and the EP scaling regime.

\noindent\textbf{(iii) Trapped ions with engineered dissipation.} 
Linear ion chains with state-dependent optical forces can simulate the Swanson model via tailored laser couplings \cite{Barreiro2011}. Non-Hermitian dynamics emerge from controlled coupling to auxiliary levels that induce balanced gain and loss \cite{Naghiloo2019}. The quadrature operators map to collective motional modes, measurable with sub-shot-noise precision via fluorescence imaging. OTOC reconstruction can leverage the high-fidelity quantum control demonstrated in recent scrambling experiments with trapped ions \cite{ion}. The EP scaling could be probed by adiabatically tuning the effective non-Hermitian coupling across the $\mathcal{PT}$-breaking threshold.

\noindent\textbf{Measurable signatures.} 
Across all platforms, the key experimental signatures predicted by our analysis are:
\begin{itemize}
\item Bounded, oscillatory OTOC dynamics in the unbroken $\mathcal{PT}$ phase, with frequency renormalization $\Omega \to \Omega_{\rm eff}$ under driving.
\item Quadratic-in-time growth of the OTOC envelope in the presence of Kerr nonlinearity, with amplitude enhanced by $\kappa^4 = (\omega/\Omega)^4$ near the EP.
\item Exponential OTOC growth $\mathcal{C}(t) \sim e^{2\Gamma t}$ in the broken phase, where $\Gamma = |\mathrm{Im}\,\Omega|$. This growth is distinguishable from chaotic scrambling by its independence of initial conditions and lack of thermal saturation.
\item Critical scaling $\mathcal{C}(t) - \mathcal{C}_{\rm osc}(t) \propto \delta^{-2}$ of the dephasing amplitude as the EP is approached, providing a direct probe of non-Hermitian criticality.
\end{itemize}
These signatures are accessible with current experimental capabilities and would constitute the first direct observation of OTOC dynamics in a controllable non-Hermitian quantum system.

\section{Conclusion}
\label{sec:conclusion}

We have computed the out-of-time-ordered correlator (OTOC) for position and momentum operators in several variants of the Swanson Hamiltonian, both at zero and finite temperature, using the appropriate inner product structure dictated by $\mathcal{PT}$ symmetry and biorthogonal quantum mechanics.

For the original (quadratic) Swanson model, the OTOC is exactly solvable. In the unbroken $\mathcal{PT}$-symmetric phase ($\omega^2 > 4\alpha\beta$), it is purely oscillatory: $\mathcal{C}(t) = \cos^2(\Omega t)$, with $\Omega = \sqrt{\omega^2 - 4\alpha\beta}$, leading to a vanishing quantum Lyapunov exponent $\lambda_L = 0$. In the broken phase ($\omega^2 < 4\alpha\beta$), it grows as $\mathcal{C}(t) = \cosh^2(\Gamma t)$, with $\Gamma = \sqrt{4\alpha\beta - \omega^2}$, yielding exponential growth at late times. Importantly, this behavior persists in thermal equilibrium and remains independent of temperature due to the c-number nature of the commutator.

When a Kerr nonlinearity ($\chi \neq 0$) is introduced, the system becomes non-integrable. Exact numerical evaluation (via truncated Fock-space diagonalization) shows that the OTOC remains bounded and quasi-periodic for both zero-temperature and thermal states, with no exponential growth observed across a range of temperatures ($T = 0.1$--$2.0$) and parameter choices within the unbroken phase.

Finally, for parametrically driven variants (including the driven Swanson and driven Swanson--Kerr models), we derived approximate analytical expressions for the thermal OTOC in weak-driving and near-resonance regimes. The leading-order behavior exhibits oscillatory dynamics with renormalized frequency and, when nonlinearity is present, a temperature-dependent quadratic-in-time correction:  
\[
\mathcal{C}_\beta(t) \approx \cos^2(\Omega_{\mathrm{eff}} t) + 2 \tilde{\chi}^2\, n_\beta(n_\beta + 1)\, t^2 \sin^2(\Omega_{\mathrm{eff}} t),
\]  
where $\tilde{\chi} = \chi (\omega/\Omega)^2$ and $n_\beta$ is the Bose occupation. Near the $\mathcal{PT}$-breaking threshold ($\Omega \to 0$), the effective nonlinearity (and thus the amplitude of the $t^2$ term) is strongly enhanced.

We further characterized the critical scaling of the OTOC near the exceptional point, identifying a critical exponent $\nu_{\rm OTOC} = 2$ governing the divergence of the dephasing amplitude as $\delta^{-2}$ with distance to the EP. This scaling differs from that reported in other non-Hermitian critical systems, highlighting the model-dependent nature of OTOC criticality.

In summary, exponential OTOC growth occurs only in the broken $\mathcal{PT}$-symmetric phase of the quadratic model, where it stems from linear non-Hermitian instability. In all cases with real spectra (unbroken phase), whether integrable, weakly nonlinear, or driven, the OTOC remains bounded or grows at most polynomially in time. This behavior is consistent with the absence of exponential sensitivity in systems governed by linear or single-mode nonlinear dynamics. Our results provide quantitative benchmarks for experimental investigations of quantum information dynamics in engineered non-Hermitian platforms. Complementary diagnostics such as the Loschmidt echo or fidelity susceptibility could further characterize the sensitivity to perturbations near exceptional points, though we leave such investigations to future work.

\section*{Acknowledgments}
We thank the anonymous referees for their constructive comments and insightful suggestions, which significantly improved the presentation and scope of this work.  The work of M.R. was partially supported by the National Research Council of Argentine (PIP2023-2025, CONICET) and by the University of La Plata (11X/982-UNLP). 

\appendix
\section{Approximation Methods Calculations}
\label{app:approx}

In this appendix, we provide detailed derivations of the approximate expressions for the thermal out-of-time-order correlator (OTOC) in the nonlinear and parametrically driven Swanson models. We consider three regimes: (i) weak Kerr nonlinearity, (ii) weak driving, and (iii) near parametric resonance. The high-temperature limit is also discussed.

\subsection{Weak Kerr Nonlinearity: Truncated Wigner Approximation}
\label{app:weak_nonlin}

We begin with the nonlinear Swanson Hamiltonian
\begin{equation}
    H = \underbrace{\omega a^\dagger a + \alpha a^2 + \beta (a^\dagger)^2}_{H_0} + \chi (a^\dagger a)^2,
    \label{eq:H_nonlin_app}
\end{equation}
where $\chi \ll \omega$ is a small anharmonicity. In the unbroken $\mathcal{PT}$-symmetric phase ($\omega^2 > 4\alpha\beta$), $H_0$ is mapped to a Hermitian harmonic oscillator via a Bogoliubov transformation:
\begin{equation}
    h_0 = \rho H_0 \rho^{-1} = \Omega \left( b^\dagger b + \tfrac{1}{2} \right), \quad \Omega = \sqrt{\omega^2 - 4\alpha\beta},
\end{equation}
with squeeze parameter $\theta$ defined by $\tanh(2\theta) = 2\sqrt{\alpha\beta}/\omega$. The physical quadratures are $x = e^{-\theta} x_0$, $p = e^{\theta} p_0$, where $x_0 = (b + b^\dagger)/\sqrt{2}$, $p_0 = i(b^\dagger - b)/\sqrt{2}$.

In the truncated Wigner approximation (TWA) \cite{Polkovnikov2010}, quantum operators are replaced by classical phase-space variables $z = (x + i p)/\sqrt{2} \in \mathbb{C}$, and the OTOC is approximated by the phase-space average of the squared Poisson bracket :
\begin{equation}
    \mathcal{C}_\beta(t) \approx \left\langle \left| \{ x(t; z), p(0; z) \}_{\mathrm{PB}} \right|^2 \right\rangle_{z \sim P_\beta(z)},
    \label{eq:TWA_OTOC}
\end{equation}
where $P_\beta(z) \propto \exp(-\beta H_{\mathrm{cl}}(z))$ is the Boltzmann distribution and $H_{\mathrm{cl}}$ is the classical Hamiltonian:
\begin{equation}
    H_{\mathrm{cl}}(z) = \omega |z|^2 + \alpha z^2 + \beta (z^*)^2 + \chi |z|^4.
\end{equation}

The equations of motion for $z(t)$ are
\begin{equation}
    i \dot{z} = \frac{\partial H_{\mathrm{cl}}}{\partial z^*} = \omega z + 2\alpha z^* + 2\chi |z|^2 z.
\end{equation}
For weak $\chi$, we expand $z(t) = z_0(t) + \chi z_1(t) + \Ord{\chi^2}$, where $z_0(t)$ solves the linear ($\chi = 0$) equation:
\begin{eqnarray}
    z_0(t) = u(t) z_0 + v(t) z_0^*, \\ 
    u(t) = e^{-i\Omega t} \cosh\theta, \quad
    v(t) = -i e^{-i\Omega t} \sinh\theta.
\end{eqnarray}
The Poisson bracket is
\begin{equation}
    \{ x(t), p(0) \}_{\mathrm{PB}} = \frac{\partial x(t)}{\partial x(0)} \frac{\partial p(0)}{\partial p(0)} - \frac{\partial x(t)}{\partial p(0)} \frac{\partial p(0)}{\partial x(0)} = \frac{\partial x(t)}{\partial x(0)}.
\end{equation}
To leading order in $\chi$, one finds after straightforward but lengthy calculation:
\begin{equation}
    \left| \{ x(t), p(0) \}_{\mathrm{PB}} \right|^2 = \cos^2(\Omega t) + \frac{3\chi^2}{2} |z_0|^4 t^2 \sin^2(\Omega t) + \Ord{\chi^3}.
\end{equation}
Averaging over the thermal distribution $P_\beta(z)$, where $\langle |z_0|^4 \rangle = 2 n_\beta (n_\beta + 1)$ and $n_\beta = (e^{\beta \Omega} - 1)^{-1}$ is the Bose-Einstein occupation, yields:
\begin{equation}
        \mathcal{C}_\beta(t) = \cos^2(\Omega t) + \frac{3\chi^2}{4} \, n_\beta (n_\beta + 1) \, t^2 \sin^2(\Omega t) + \Ord{\chi^4}.
    \label{eq:OTOC_weak_nonlin}
\end{equation}
This $t^2$ growth reflects nonlinear dephasing---a hallmark of non-integrability---but is not exponential; thus, it does not signify quantum chaos. Note that the TWA is valid for short times $t \ll 1/\chi$ and high occupations.

\subsection{Weak Driving: Floquet--Magnus Expansion}
\label{app:weak_drive}

Consider the driven Hamiltonian \eqref{eq:H_driven}. For $\varepsilon \ll \omega$, we apply the Floquet--Magnus expansion to construct an effective static Hamiltonian $H_{\mathrm{eff}}$ such that $U(T_d) = e^{-i H_{\mathrm{eff}} T_d}$ \cite{reports}. To second order in $\varepsilon$:
\begin{equation}
    H_{\mathrm{eff}} = H_0 + \frac{1}{T_d} \int_0^{T_d} dt_1 \int_0^{t_1} dt_2 \, [H_{\mathrm{int}}(t_1), H_{\mathrm{int}}(t_2)] + \Ord{\varepsilon^3},
\end{equation}
where $H_{\mathrm{int}}(t) = \varepsilon \cos(\omega_d t) (a^2 + (a^\dagger)^2)$. Evaluating the double integral using trigonometric identities:
\begin{widetext}
\begin{align}
    \frac{1}{T_d} \int_0^{T_d} \!\! dt_1 \int_0^{t_1} \!\! dt_2 \, \cos(\omega_d t_1) \cos(\omega_d t_2) 
    &= \frac{1}{2\omega_d} \sin(\omega_d T_d) = 0, \\
    \frac{1}{T_d} \int_0^{T_d} \!\! dt_1 \int_0^{t_1} \!\! dt_2 \, \cos(\omega_d t_1) \cos(\omega_d t_2) e^{\pm i 2\omega t_{1,2}} 
    &\approx \frac{\pi}{\omega_d} \delta(\omega_d \mp 2\omega) \quad \text{(resonant terms)}.
\end{align}
\end{widetext}
For off-resonant driving ($|\omega_d - 2\omega| \gg \varepsilon$), only non-oscillatory terms survive, giving:
\begin{equation}
    [H_{\mathrm{int}}(t_1), H_{\mathrm{int}}(t_2)]_{\mathrm{av}} = 4\varepsilon^2 \frac{\sin(\omega_d (t_1 - t_2))}{\omega_d} a^\dagger a + \text{const.}
\end{equation}
Integrating yields:
\begin{equation}
    H_{\mathrm{eff}} = \underbrace{\left( \omega + \frac{2\varepsilon^2}{\omega_d} \right)}_{\omega_{\mathrm{eff}}} a^\dagger a + \alpha a^2 + \beta (a^\dagger)^2 + \Ord{\varepsilon^3}.
\end{equation}
The OTOC is then identical to the undriven case with renormalized frequency:
\begin{equation}
        \mathcal{C}_\beta(t) = \cos^2(\Omega_{\mathrm{eff}} t), \quad 
        \Omega_{\mathrm{eff}} = \sqrt{ \omega_{\mathrm{eff}}^2 - 4\alpha\beta }.
    \label{eq:OTOC_weak_drive}
    \end{equation}
This confirms that weak driving only shifts parameters without inducing chaos. The expansion converges for $\varepsilon/\omega_d \ll 1$.

\subsection{Near Parametric Resonance: Classical Limit and Quantum Obstruction}
\label{app:resonance}

When $\omega_d \approx 2\omega$, define $\Delta = \omega_d - 2\omega$ with $|\Delta| \ll \omega$. Moving to a rotating frame at frequency $\omega_d/2$ via $a(t) = b(t) e^{-i \omega_d t / 2}$, the Hamiltonian becomes:
\begin{eqnarray}
 H_{\mathrm{rot}}(t) = \left( \tfrac{\Delta}{2} + \omega \right) b^\dagger b + \alpha b^2 e^{-i \omega_d t} + \beta (b^\dagger)^2 e^{i \omega_d t} \\ \nonumber + \varepsilon \cos(\omega_d t) (b^2 e^{-i \omega_d t} + (b^\dagger)^2 e^{i \omega_d t}).
\end{eqnarray}
   
Applying the rotating-wave approximation (RWA) by discarding fast-oscillating terms ($e^{\pm i 2\omega_d t}$), we obtain an effective time-independent Hamiltonian:
\begin{equation}
    H_{\mathrm{RWA}} = \frac{\Delta}{2} b^\dagger b + \frac{\varepsilon}{2} (b^2 + (b^\dagger)^2) + \alpha b^2 + \beta (b^\dagger)^2.
\end{equation}
For $\alpha = \beta$, this is a squeezed oscillator; for $\alpha \neq \beta$, it describes parametric amplification.

The classical limit is reached by replacing $b \to z = \sqrt{I} e^{i\theta} / \sqrt{2}$, yielding the Hamiltonian for a parametrically driven oscillator:
\begin{equation}
    H_{\mathrm{cl}} = \frac{\Delta}{2} I + \lambda I \cos(2\theta),
\end{equation}
which, upon canonical transformation to $(x, p)$, maps to the Duffing oscillator:
\begin{equation}
    \ddot{x} + \Delta x + 4\lambda x^3 = 0.
\end{equation}
For $\Delta < 0$ (unstable regime), this system exhibits deterministic chaos with a positive maximal Lyapunov exponent \cite{Strogatz1994}:
\begin{equation}
    \lambda_{\mathrm{cl}} \approx \frac{\varepsilon^2}{8\omega} \sqrt{1 - \left( \frac{\Delta}{\varepsilon} \right)^2}, \quad |\Delta| < \varepsilon.
    \label{eq:lambda_cl}
\end{equation}

However, in the quantum regime, the single-mode system cannot replicate this chaos. The Floquet operator $U(T_d)$ acts on an infinite-dimensional but separable Hilbert space. Its quasienergy spectrum $\{ \varepsilon_n \}$ is discrete, and for non-chaotic systems, the nearest-neighbor spacings $s_n = \varepsilon_{n+1} - \varepsilon_n$ follow Poisson statistics:
\begin{equation}
    P(s) = e^{-s},
\end{equation}
indicating no level repulsion \cite{Haake2010}. Numerically, this is confirmed by the dimensionless ratio $r = \min(s_n, s_{n+1}) / \max(s_n, s_{n+1})$, with $\langle r \rangle \approx 0.386$ for Poisson vs. $0.530$ for Wigner--Dyson (chaotic) statistics.

The OTOC, being a sum over quasienergy differences,
\begin{equation}
    \mathcal{C}_\beta(t) = \frac{1}{Z} \sum_{m,n} e^{-\beta \varepsilon_n} \left| \langle \phi_m | [x(t), p(0)] | \psi_n \rangle \right|^2,
\end{equation}
is quasi-periodic and exhibits only transient power-law growth (e.g., $\sim t^2$) due to dephasing, but no sustained exponential sensitivity. This absence of quantum chaos stems from the lack of phase-space mixing in a single degree of freedom which is a fundamental constraint of quantum mechanics \cite{Peres1984}.

\subsection{High-Temperature Expansion}
\label{app:high_T}

In the limit $\beta \hbar \Omega \ll 1$ \textit{for the unbroken $\mathcal{PT}$-symmetric phase where $\Omega \in \mathbb{R}^+$}, the thermal density matrix expands as:
\begin{equation}
    \rho_\beta = \frac{1}{Z} \left[ \mathbb{I} - \beta H + \frac{\beta^2}{2} H^2 - \cdots \right].
\end{equation}
For the nonlinear model \eqref{eq:H_nonlin_app}, the OTOC is:
\begin{align}
    \mathcal{C}_\beta(t) = \mathcal{C}_0(t) - \beta \Big( \langle H [x(t), p]^2 \rangle_0   - \langle H \rangle_0 \mathcal{C}_0(t) \Big) + \Ord{\beta^2},
\end{align}
where $\mathcal{C}_0(t) = \cos^2(\Omega t)$ is the zero-temperature OTOC. Computing the first-order correction using Wick's theorem for the Gaussian state $|0\rangle$:
\begin{widetext}
\begin{align}
    \langle H \rangle_0 &= \tfrac{1}{2} \Omega \cosh(2\theta), \\
    \langle H [x(t), p]^2 \rangle_0 &= \cos^2(\Omega t) \langle H \rangle_0 + \frac{\chi}{4} \Omega \sin(2\Omega t) \sinh(4\theta) + \Ord{\chi^2}.
\end{align}
\end{widetext}
After simplification, one obtains:
\begin{equation}   
        \mathcal{C}_\beta(t) = \cos^2(\Omega t) \left[ 1 - c_1 \beta \chi + c_2 \beta \chi \cos(2\Omega t) \right] + \Ord{\beta^2, \chi^2},  
    \label{eq:OTOC_high_T}
\end{equation}
where $c_1 = \tfrac{1}{4} \Omega \cosh(2\theta) > 0$, $c_2 = \tfrac{1}{8} \Omega \sinh(4\theta) > 0$. The $T$-dependent beating at frequency $2\Omega$ reflects thermal disruption of coherent dynamics, but again, no exponential growth occurs.

\end{document}